\documentclass[aps,prm,reprint,longbibliography,citeautoscript,%
superscriptaddress,showpacs,showkeys]{revtex4-1}
\usepackage{graphicx,epsfig,float}
\usepackage{bm}
\usepackage{amssymb}
\usepackage{amsmath}
\usepackage[varg]{txfonts}% times in math
\allowdisplaybreaks
\renewcommand{\vec}[1]{\bm{\mathrm{#1}}}     %вектор --- жирный шрифт
\makeatletter
\let\oldsqrt\sqrt
\renewcommand{\sqrt}[1]{\!\oldsqrt{#1}}
\mathchardef\originalS\mathcode`S
\begingroup\lccode`~=`S
\lowercase{\endgroup\def~}{\originalS\kern-1.2pt}
\mathcode`S="8000
\makeatother

\begin{document}
\title{Restriction of macroscopic structural superlubricity due to structure relaxation\\ by the example of twisted graphene bilayer}

\author{Alexander S. Minkin}
\email{amink@mail.ru}
\affiliation{Keldysh Institute of Applied Mathematics Russian Academy of Sciences,
4 Miusskaya sq., Moscow, 125047, Russia}

\author{Irina V. Lebedeva}
\email{liv\_ira@hotmail.com}
\affiliation{CIC nanoGUNE BRTA, San Sebasti\'an 20018, Spain}
\affiliation{Simune Atomistics, Avenida de Tolosa 76, San Sebasti\'an 20018, Spain}

\author{Andrey M. Popov}
\email{popov-isan@mail.ru}
\affiliation{Institute of Spectroscopy of Russian Academy of Sciences, Fizicheskaya str.~5, Troitsk, Moscow 108840, Russia}

\author{Sergey A. Vyrko}
\email{vyrko@bsu.by}
\affiliation{Physics Department, Belarusian State University, Nezavisimosti Ave.~4, Minsk 220030, Belarus}

\author{Nikolai A. Poklonski}
\email{poklonski@bsu.by}
\affiliation{Physics Department, Belarusian State University, Nezavisimosti Ave.~4, Minsk 220030, Belarus}

\author{Yurii E. Lozovik}
\affiliation{Institute of Spectroscopy of Russian Academy of Sciences, Fizicheskaya str.~5, Troitsk, Moscow 108840, Russia}

\begin{abstract}
The effect of structure relaxation on the potential energy surface (PES) of interlayer interaction of twisted graphene bilayer is studied for a set of commensurate moir\'e systems using the registry-dependent empirical potential of Kolmogorov and Crespi. It is found that the influence of structure relaxation on the amplitude of PES corrugations (determining static friction) depends on the unit cell size (or related twist angle) of the moir\'e system. For moir\'e systems with the smallest unit cells, the amplitudes of PES corrugations calculated with and without account of structure relaxation are approximately the same. However, for large unit cell sizes, the structure relaxation can lead to an increase of PES corrugations by orders of magnitude. This means that structure relaxation can provide the main contribution into the static friction of a superlubric system under certain conditions (such as the contact size and twist angle). Moreover, the change of the PES type because of structure relaxation from a trigonal lattice of maxima to a trigonal lattice of minima is observed for the systems with the moir\'e patterns (5,1) and (5,3). Based on the results obtained, possible crossovers between static friction modes taking place upon changing the twist angle in a macroscopic superlubric system consisting of identical layers are discussed. Additionally it is shown that the PES for relaxed structures can still be approximated by the first Fourier harmonics compatible with symmetries of twisted layers analogously to the PES for rigid layers.
\end{abstract}

%Use showkeys class option if keyword display desired

\maketitle

\section{Introduction}

Incommensurability makes possible relative motion of surfaces with extremely low friction \cite{Hirano1990, Hirano1991}. This phenomenon, referred to as structural superlubricity, has been actively studied in recent years for systems based on graphene, hexagonal boron nitride and other two-dimensional (2D) materials (see, for example,~\cite{Hod2018} for a review). In the present paper, we study the influence of structure relaxation on the area contribution into static friction of macroscopic superlubric systems by the example of twisted graphene bilayers.

First structural superlubricity was detected for graphene flakes at a probe tip sliding on graphite \cite{Verhoeven2004, Dienwiebel2005, Filippov2008}. Since then a large number of atomistic simulations have been performed to get insight into tribological properties of superlubric systems with a small contact area in which the rim or edge contribution to static friction is dominant, e.g., a graphene flake on graphene or a graphite surface \cite{Verhoeven2004, Filippov2008, Xu2013, Koren2016, Bonelli2009, vanWijk2013, Guo2007, Shibuta2011, Yoon2014, Zhang2015a, Zhang2018, Wang2019a, Zhang2021, Zhang2022, Bai2022, Bai2022a, Tang2023, Yan2024}. Note that in the systems with the contact between the same 2D materials without tension, the superlubricity can be lost via the rotation of the layers to the commensurate relative orientation \cite{Hirano1990, Filippov2008, Xu2013, Bonelli2009, vanWijk2013, Guo2007, Shibuta2011, Zhang2015a}.

Robust superlubricity has been recently achieved for systems with a lattice mismatch such as heterostructures composed of layers of different 2D materials \cite{Song2018} or layers of the same 2D material under different tension applied \cite{Wang2019, Androulidakisl2020}. For such systems, relative rotation of the layers to the commensurate configuration accompanied by the loss of superlubricity is not possible. It has been also proposed that macroscopic robust superlubricity might be observed for superlubric systems with commensurate moir\'e patterns due to the existence of a barrier for rotation of the layers to the fully incommensurate relative orientation \cite{Minkin2023}. Recent experiments \cite{Dietzel2013, Koren2015, Kawai2016, Qu2020} and theoretical studies \cite{Koren2016, Wijn2012, Muser2001, Mandelli2017, Zhang2021} suggest that the static friction force per unit area decreases for superlubric systems upon increasing the contact area. Therefore, it is important to analyze possible sources of static friction at macroscale incommensurate interfaces or in other words the reasons for restriction of macroscopic structural superlubricity \cite{Liu2012, Mandelli2017, Hod2018, Koren2016, Minkin2021, Minkin2022, Minkin2023, Bai2022, Feng2022, Bai2023}.

Up to now the following contributions to static friction in 2D systems with macroscopic robust superlubricity have been studied: 1) incomplete cancelation of static friction forces within full unit cells of a commensurate moir\'e pattern (area contribution) \cite{Koren2016, Minkin2023}, 2) forces coming from partial unit cells in the rim area of one of the layers (rim contribution), \cite{Koren2016, Qu2020} 3) friction induced by grain boundaries \cite{Liu2012} and atomic-scale defects \cite{Liu2012,Minkin2021,Minkin2022}, 4) motion of domain walls in superstructures formed upon relaxation of moir\'e patterns with spatial periods that are much greater than the domain wall width \cite{Mandelli2017, Hod2018, Feng2022}, and 5) deformation of 2D layers related with interaction with a substrate \cite{Bai2023}. The contribution of atomic-scale defects into static friction has been also investigated for superlubric relative sliding and rotation of nanotube walls \cite{Belikov2004, Shibuta2011, Zhang2013a}. It is also worth mentioning studies of atomistic mechanisms of dynamic friction in macroscopic 2D superlubric systems \cite{Song2018, Mandelli2017, Wang2019b, Brilliantov2023}.

Previously it has been reported that structure relaxation strongly affects static friction in superlubric systems with a finite contact area in which the edge or rim area contribution into static friction is dominant \cite{Bonelli2009, Zhang2015a, Wang2019, Zhang2018, Zhang2022}. However, tribological properties of infinite superlubric systems related to cancelation of static friction forces within complete unit cells of commensurate moir\'e patterns of graphene bilayer \cite{Koren2016, Minkin2021, Minkin2022, Minkin2023} and double-walled carbon nanotubes \cite{Kolmogorov2000, Belikov2004, Bichoutskaia2006} have been considered only for rigid structures of the layers. Here we propose that structure relaxation can lead to a significant increase of the area contribution to static friction and, thus, can restrict the superlubricity in macroscopic 2D superlubric systems. To prove this statement, we study the effect of structure relaxation on static friction by the example of infinite twisted graphene bilayers with commensurate moir\'e patterns.

The static friction of 2D systems is determined by the potential energy surface (PES) that is the dependence of the potential energy on the relative in-plane displacement of 2D layers. Such PESs are calculated here with and without account of structure relaxation using the registry-dependent Kolmogorov--Crespi potential \cite{Kolmogorov2005}. We investigate the influence of structure relaxation on the amplitude of PES corrugations, i.e.~the difference between maximum and minimum values of the energy, and the dependence of this effect on the size of the unit cell of commensurate moir\'e patterns and the related twist angle. The calculations performed allow us getting an insight into crossovers between static friction modes that occur when the twist angle is changed.

The PESs of interlayer interaction for layered 2D materials with commensurate layers aligned in the same or opposite directions are universally described by the first spatial Fourier harmonics compatible with symmetries of the both layers \cite{Lebedev2020}. This hypothesis has been confirmed by PES calculations for a wide set of 2D materials \cite{Ershova2010, Lebedeva2011, Popov2012, Lebedeva2012, Zhou2015, Reguzzoni2012, Lebedev2016, Lebedev2020}, 2D heterostructures \cite{Jung2015, Kumar2015, Lebedev2017} and commensurate double-walled carbon nanotubes \cite{Vucovic2003, Belikov2004, Bichoutskaia2005, Bichoutskaia2009, Popov2009, Popov2012a}. Moreover, we showed recently that the first Fourier harmonics are also sufficient for description of the PES for twisted layers forming an infinite commensurate moir\'e pattern \cite{Minkin2023}. Here we demonstrate that structure relaxation does not affect the accuracy of this approximation even in the cases when it causes changes in the amplitude of PES corrugations by orders of magnitude.

The paper is organized as follows. In Sec.~II, we describe the atomic model of twisted graphene bilayer and computational methods. In Sec.~III, we present the results on the influence of structure relaxation on superlubricity and PES approximation by the first Fourier harmonics. Sec.~IV is devoted to the discussion and conclusion.

\section{Methodology}

\begin{figure}
   \centering
 \includegraphics{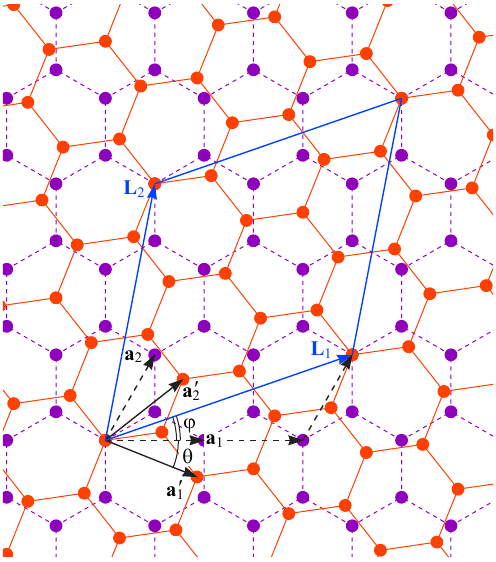}
   \caption{(Color online) A scheme of the commensurate moir\'e pattern (2,1) of a twisted graphene bilayer. Lattice vectors $\vec{a}_1$ and $\vec{a}_2$ of the bottom graphene layer and $\vec{a}_1'$ and $\vec{a}_2'$ of the top layer, lattice vectors $\vec{L}_1$ and $\vec{L}_2$ of the commensurate moir\'e pattern, angle $\varphi$ between the lattice vector $\vec{a}_1$ of the bottom layer and lattice vector $\vec{L}_1$ of the commensurate moir\'e pattern, and twist angle $\theta$ of relative rotation of the graphene layers are indicated.}
   \label{fig:01}
\end{figure}

\subsection{Model of infinite twisted graphene bilayer}

Modeling of macroscopic superlubricity using atomistic methods is not an easy task. On the one hand, atomistic simulations of infinite incommensurate superlubric systems are not possible since such systems do not satisfy periodic boundary conditions. On the other hand, structure relaxation related to interlayer interaction occurs near edges of 2D layers in a different manner than far from the edges. The region close to the edge of the width of about 10 nm is reconstructed \cite{Lebedeva2017}. The account of structure relaxation in atomistic simulations implies an increase of the computational time by orders of magnitude. All these factors make it difficult to calculate the whole PES with account of structure relaxation for incommensurate superlubric systems with edges and sizes that correspond to macroscopic superlubricity even using empirical classical potentials. Thus, a twisted graphene bilayer with a commensurate moir\'e pattern (that satisfies the periodic boundary conditions) is a preferred model for our atomistic studies of macroscopic superlubricity. This model has been already used previously to investigate the restriction of structural superlubricity by atomic-scale defects \cite{Minkin2021, Minkin2022}.

A commensurate moir\'e pattern $(n_1,n_2)$ of a twisted graphene bilayer is defined by coprime indexes $n_1$ and $n_2$ \cite{Mele2012}. The commensurate moir\'e pattern (2,1) is shown in Fig.~1, lattice vectors of graphene $\vec{a}_1$ and $\vec{a}_2$ of the bottom layer and $\vec{a}_1'$ and $\vec{a}_2'$ of the top layer, and the moir\'e pattern unit cell are indicated. For each commensurate moir\'e pattern $(n_1,n_2)$ there is a twin pattern $(n_1',n_2')$ with the same size of the unit cell \cite{Mele2012, Campanera2007}. Considering pairs of indexes of twin moir\'e patterns, $(n_1-n_2)/3$ is not integer for the smaller indexes and integer for the greater ones. Pairs of twin moir\'e patterns with the same size of the pattern unit cell and with different symmetry have been considered as different in the original work \cite{Mele2012}. Lately it has been shown that twin commensurate moir\'e patterns can be obtained one from the other by a translation of one of the layers in the layer plane \cite{Minkin2023}. Therefore, twin commensurate moir\'e patterns have the same PES and only patterns with smaller indexes [$(n_1-n_2)/3$ is not integer] are considered here.

In this case, the unit cell of the commensurate moir\'e pattern $(n_1,n_2)$ is determined by the lattice vectors $\vec{L}_1$ and $\vec{L}_2$ as
\[
   \vec{L}_1 = n_1\vec{a}_1 + n_2\vec{a}_2,\quad
   \vec{L}_2 = -n_2\vec{a}_1 + (n_1+n_2)\vec{a}_2,
\]
where the length of the lattice vectors is $L = |\vec{L}_1| = |\vec{L}_2| = a\sqrt{n_1^2 + n_1n_2 + n_2^2}$. The twist angle $\theta$ of relative rotation of graphene layers (that is the angle between the vectors $\vec{a}_1$ and $\vec{a}_1'$) is determined by
\[
   \cos\theta = \frac{n_1^2 + 4n_1n_2 + n_2^2}{2(n_1^2 + n_1n_2 + n_2^2)}.
\]

The angle $\varphi$ between the lattice vector $\vec{a}_1$ of the bottom graphene layer and the lattice vector of the commensurate moir\'e pattern $\vec{L}_1$ is given by
\[
   \varphi = 30^\circ - \frac{\theta}{2}.
\]

The angles $\theta$ and $\varphi$ are indicated in Fig.~1. The area of the unit cell of the moir\'e pattern $(n_1,n_2)$ can be computed as
\[
   S = S_gN_c = \frac{\sqrt{3}a^2(n_1^2 + n_1n_2 + n_2^2)}{2},
\]
where $N_c = n_1^2 + n_1n_2 + n_2^2$ is the number of graphene unit cells per the moir\'e pattern unit cell, $S_g = \sqrt{3}a^2/2$ is the area of the graphene unit cell, and $a = |\vec{a}_1| = |\vec{a}_2|$ is the lattice constant of graphene.

\subsection{Computational details}

\begin{table*}
\caption{Angle $\theta$ of relative rotation of graphene layers, number $N_c$ of graphene unit cells per the moir\'e pattern unit cell, lateral simulation cell size in units of moir\'e pattern unit cells, calculated amplitude $\Delta U_\mathrm{max}$ of PES corrugations and barrier $\Delta U_\mathrm{b}$ to relative sliding of the layers (both in $\mu$eV per atom of the top layer) for commensurate moir\'e patterns $(n_1,n_2)$ with rigid graphene layers, layers relaxed with constraints on in-plane positions of all atoms and only two atoms in the simulation cell.}
\renewcommand{\arraystretch}{1.2}
\setlength{\tabcolsep}{8pt}
%\resizebox{\columnwidth}{!}{
\begin{tabular}{*{10}{c}}
\hline
\hline
 &  &  &  & \multicolumn{2}{c}{rigid layers} & \multicolumn{2}{c}{out-of-plane relaxation} & \multicolumn{2}{c}{relaxation with two atoms constrained} \\
$(n_1,n_2)$ & $\theta$ ($^\circ$) & $N_c$ & cell size & $\Delta U_\mathrm{max}$ & $\Delta U_\mathrm{b}$ & $\Delta U_\mathrm{max}$ & $\Delta U_\mathrm{b}$ & $\Delta U_\mathrm{max}$ & $\Delta U_\mathrm{b}$\\\hline
(2,1) & 21.787 & 7  & $6\times6$ & 87.7 & 75.4 & 94.6 & 81.9 & 93.8 & 81.1 \\
(3,1) & 32.204 & 13 & $5\times5$ & 20.6 & 2.29 & 20.4 & 2.26 & 20.4 & 2.27 \\
(3,2) & 13.174 & 19 & $4\times4$ & 1.87 & 0.208 & 4.83 & 0.538 & 4.20 & 0.467 \\
(5,1) & 42.103 & 31 & $12\times12$ & $4.72{\cdot}10^{-3}$ & $5.25{\cdot}10^{-4}$ & $8.68{\cdot}10^{-2}$ & $7.72{\cdot}10^{-2}$ &  $5.79{\cdot}10^{-2}$ & $5.15{\cdot}10^{-2}$ \\
(5,3) & 16.426 & 49 & $10\times10$ &  $<$8.17${\cdot}10^{-7}$ &  & $1.20{\cdot}10^{-2}$ & 1.06$\cdot10^{-2}$ & $7.45{\cdot}10^{-3}$ & 6.62$\cdot10^{-3}$ \\
\hline
\hline
\end{tabular}
%}
\label{table:01}
\end{table*}

\begin{table*}
   \caption{Calculated optimal interlayer distance $d_\mathrm{min}$ (in \AA) for commensurate moir\'e patterns $(n_1,n_2)$ with rigid graphene layers, layers relaxed with constraints on in-plane positions of all atoms and only two atoms in the simulation cell, difference $\Delta d$ (in \AA) between maximum and minimum average interlayer distances for relaxed bilayers, and minimum and maximum corrugations $b$ (in \AA) of the layer plane for relaxed bilayers (difference between maximum and minimum $z$-coordinate of atoms within one layer).}
   \renewcommand{\arraystretch}{1.2}
   \setlength{\tabcolsep}{8pt}
   %\resizebox{\columnwidth}{!}{
   \begin{tabular}{*{10}{c}}
      \hline
      \hline
       & rigid layers & \multicolumn{4}{c}{out-of-plane relaxation} & \multicolumn{4}{c}{relaxation with two atoms constrained}\\
      $(n_1,n_2)$ & $d_\mathrm{min}$ & $d_\mathrm{min}$ & $\Delta d_\mathrm{max}$ & $b_\mathrm{min}$ & $b_\mathrm{max}$ & $d_\mathrm{min}$ & $\Delta d_\mathrm{max}$ & $b_\mathrm{min}$ & $b_\mathrm{max}$ \\\hline
      (2,1) & 3.458748 & 3.456689 & 2.70$\cdot10^{-3}$ & 0.00557 & 0.00911 & 3.456537 & 2.67$\cdot10^{-3}$ & 0.00560 & 0.00911 \\
      (3,1) & 3.460108 & 3.458742 & 5.27$\cdot10^{-4}$ & 0.00395 & 0.00413 & 3.458601 & 5.31$\cdot10^{-4}$ & 0.00393 & 0.00477 \\
      (3,2) & 3.460268 & 3.455295 & 1.71$\cdot10^{-4}$ & 0.0313 & 0.0389 &  3.454830 & 1.40$\cdot10^{-4}$ & 0.0313 & 0.0390 \\
      (5,1) & 3.460235 & 3.457536 & $<$7.42$\cdot10^{-6}$ & 0.0152 & 0.0163 & 3.457280 & $<$3.38$\cdot10^{-6}$ & 0.0153 & 0.0165 \\
      (5,3) & 3.460236 & 3.457005 & $<$3.27$\cdot10^{-6}$ & 0.0200 & 0.0217 & 3.456702 & $<$1.74$\cdot10^{-6}$ & 0.0201 & 0.0218 \\
      \hline
      \hline
   \end{tabular}
   %}
   \label{table:02}
\end{table*}

Both the amplitude of PES corrugations and PES period are small for twisted layers and decrease fast upon increasing the size of the moir\'e pattern unit cell \cite{Xu2013, Koren2016, Minkin2022, Minkin2023}. This means that very high energy and spatial resolution are needed for studies of PESs of twisted layers. Such calculations using {\it ab initio} methods would have a huge computational cost even for moir\'e patterns with small unit cell sizes. An alternative is to use classical potentials. Although we cannot expect them to be quantitatively accurate and they might suffer from multiple local minima that can hinder getting to the global energy minimum when optimizing the structure geometry, classical potentials make possible qualitative studies of PESs for several sizes of moir\'e pattern unit cells at a moderate computational cost \cite{Minkin2023}. Following our previous work \cite{Minkin2023} devoted to  infinite superlubric systems with rigid layers, we use here the registry-dependent Kolmogorov--Crespi classical potential \cite{Kolmogorov2005} as implemented in LAMMPS \cite{Thompson2022}. (The detailed discussion on existing classical potentials for the description of interaction between graphene layers is presented in \cite{Minkin2023}.) Note that currently there are no experimental data characterizing physical properties of twisted graphene layers that can be used to verify the adequacy of existing classical potentials for the description of such systems. Thus, the aim of the present study is to determine only the qualitative influence of structure relaxation on tribological properties of superlubric systems consisting of twisted layers.

\begin{figure*}
   \centering
 \includegraphics{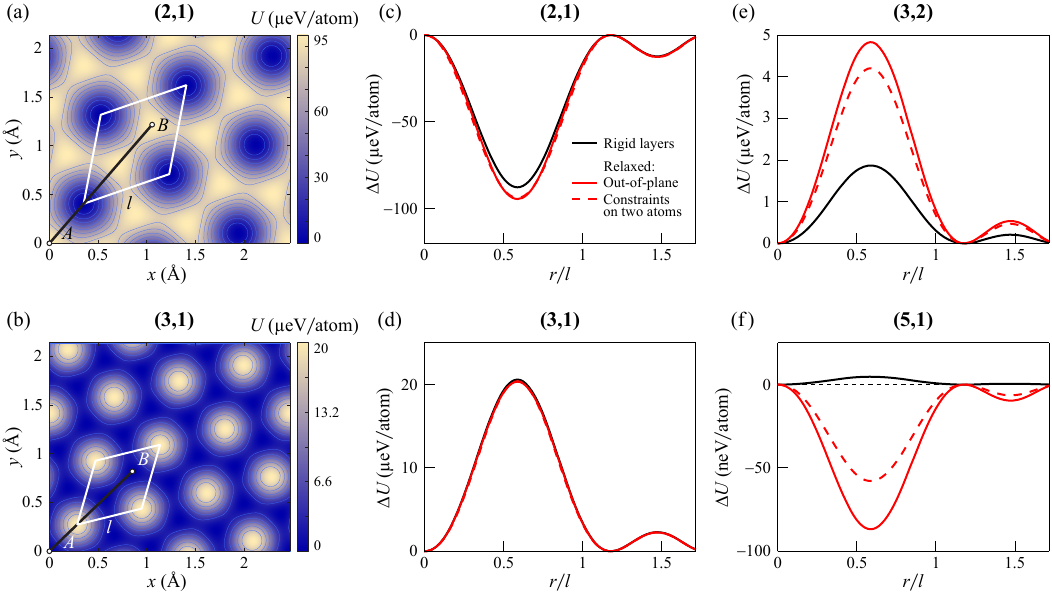}
   \caption{(Color online) (a), (b)  Potential energy $U$ of interlayer interaction (per atom of the top layer) of twisted graphene bilayers as a function of the relative displacement of the layers in the zigzag ($x$, in \AA) and armchair ($y$, in \AA) directions of the bottom layer calculated with account of out-of-plane relaxation for commensurate moir\'e patterns (a) (2,1) and (b) (3,1) corresponding to the first and second types of the potential energy surface (PES), respectively (see the text). The point $(x,y)=(0,0)$ corresponds to the stacking of the layers shown in Fig.~1. The energy is given relative to the minimum. The unit cells of the PESs are shown as white diamonds, and $l$ is the length of the PES lattice vector. Segments AB are used to calculate the PES profiles. (c)--(f) Potential energy change $\Delta U = U-U(0,0)$ as a function of the relative displacement $r/l$ of the layers along the segment AB for twisted graphene bilayers with commensurate moir\'e patterns (c) (2,1), (d) (3,1), (e) (3,2) and (f) (5,1) calculated for rigid layers (black lines), layers relaxed with constraints on in-plane positions of all atoms (red solid lines), and only two atoms in the simulation cell (red dashed lines).}
   \label{fig:02}
\end{figure*}

The amplitude of PES corrugation exceeds the accuracy of the PES calculation using the Kolmogorov--Crespi potential for rigid graphene layers only for 5 commensurate moir\'e patterns (2,1), (3,1), (3,2), (5,1), and (5,3) \cite{Minkin2023}. The calculations have been performed for these 5 commensurate moir\'e patterns under the periodic boundary conditions. The cutoff radius of the potential of 16 \AA{} was used for the moir\'e patterns (2,1), (3,1), and (3,2), and 70~\AA{} for the moir\'e patterns (5,1) and (5,3). The height of the simulation cell was 100~\AA{} for all the moir\'e patterns. The lateral sizes of the simulation cell in the units of moir\'e pattern unit cells for the considered moir\'e patterns are listed in Table~\ref{table:01}.

The interaction between atoms within the graphene layers was described using the second-generation Brenner (REBO-2002) potential \cite{Brenner2002}. The bond length of 1.42039 \AA, which is optimal for graphene according to the REBO-2002 potential, was used to build the structures of twisted layers. The PESs for rigid layers were computed at the optimal interlayer distance for each moir\'e pattern (Table~\ref{table:02}), which was obtained by the calculation of the energy dependence on the interlayer distance. The structure optimization was performed using the Polak--Ribi\`ere version \cite{Polak1969} of the conjugate gradient algorithm \cite{Press2007}. The optimization was stopped when the energy change between successive iterations divided by the energy magnitude was less than $10^{-15}$ or forces on all atoms were smaller than $10^{-15}$~eV/atom.

In order to compute the PES, it is needed to apply constraints that would keep the system away from the energy minimum. In the present paper, we consider two types of constraints that allow getting insight in the potential energy dependence on the local stacking of twisted layers. In the first case, in-plane positions of all atoms of the layers are fixed but atoms can relax out of the plane. Such a constraint reveals the contribution of out-of-plane relaxation to the PES. To estimate how in-plane displacements affect the PES, we also consider the case when in-plane positions are fixed only for one atom of the bottom layer and one atom of the top layer, while the rest of the atoms are free. A constrained atom in the bottom layer is chosen close to the simulation box vertex and the other one in the top layer close to the middle of the simulation box. We have checked that the choice of constrained atoms has a negligible effect on the PES shape and amplitude as long as the atoms are sufficiently separated [the relative changes in the PES amplitude for different choices of atom pairs vary within 0.0005\% for the moir\'e patterns (2,1) and (3,2)]. Since only two atoms per supercell are constrained, the majority of moir\'e unit cells in the supercell can relax freely.

Examples of PESs for the commensurate moir\'e patterns (2,1) and (3,1) calculated with account of out-of-plane relaxation are shown in Figs.~2a and 2b, respectively. The initial stacking of layers $(x,y) = (0,0)$ corresponds to the in-plane relative rotation of the layers by the angle $\theta$ around the vertical axis which passes through one atom in each of the layers. Two types of PESs have been found for graphene bilayers with commensurate moir\'e patterns \cite{Minkin2023}. On PESs of the first type, minima are located in vertices of a trigonal lattice, while maxima are aligned on a hexagonal lattice (Fig.~2a). The opposite takes place for PESs of the second type (Fig.~2b). The stacking with $(x,y)=(0,0)$ corresponds to a maximum for PESs of the first type and a minimum for PESs of the second type. The lattice vectors $\vec{l}_1$ and $\vec{l}_2$ of the PES of the graphene bilayer with the commensurate moir\'e pattern are directed along the lattice vectors $\vec{L}_1$ and $\vec{L}_2$ of the commensurate moir\'e pattern, respectively, and have the length $|\vec{l}_1| = |\vec{l}_2| = L/N_c$ (see Sec.~IIIB). Thus, the unit cell of each of the graphene layers corresponds to $N_c$ unit cells of the PES and the unit cell of the commensurate moir\'e pattern contains $N^2_c$ complete unit cells of the PES.

Note that the PES symmetry is related to the symmetry of the commensurate moir\'e pattern so that singular points of the PES (minima, maxima and saddle points) correspond to high-symmetry stackings of the layers. Such stackings are known without calculation of the potential energy. Because of positions of extrema on the PES being determined by the system symmetry, these positions are not changed during structure relaxation. Since the calculations of the full PESs with account of structure relaxation for all the considered moir\'e patterns would require a significant computational time, for most of them we have computed only dependences of the potential energy on the relative displacement of the layers (PES profiles) along the line segment AB connecting two equivalent extrema and passing additionally through a minimum, maximum and saddle point (Fig.~2). Such a segment has the length $\sqrt{3}l$ and is directed along the diagonal of the commensurate moir\'e pattern unit cell. The structure relaxation was performed for 102 equidistant points within the segment.

\section{Results}
\subsection{Influence of structure relaxation}

The potential energy as a function of the relative displacement $r/l$ of the layers along the segment AB for twisted graphene bilayers with commensurate moir\'e patterns (2,1), (3,1), (3,2), and (5,1) is shown in Figs.~2c-f. As discussed below in Sec.~IIIB, the PES of a graphene bilayer with a commensurate moir\'e pattern can be approximated by the first Fourier harmonics with a high accuracy. This means that the PES shape is nearly the same for different commensurate moir\'e patterns with the same type of PES and the whole PES is characterized by a single energy parameter. Since static friction of superlubric systems is determined by PES corrugations, the amplitude $\Delta U_\mathrm{max}$ of such corrugations can be used to compare the influence of structure relaxation on this physical property for different commensurate moir\'e patterns (the relation between the PES and static friction force is discussed below in Sec.~IIIB). Here we define this amplitude, which we use as a single energy parameter describing the PES and static friction, as the difference between the maximum and minimum values of the potential energy. The corresponding values of $\Delta U_\mathrm{max}$ are listed in Table~\ref{table:01}.

As for the moir\'e pattern (5,3) with the largest considered unit cell, the PES corrugations for rigid layers in this case are rather small and the PES shows significant fluctuations related to calculation errors. Therefore, we can only provide an upper estimate of $\Delta U_\mathrm{max}$ (Table~\ref{table:01}). The account of structure relaxation, however, results in a significant increase of the PES corrugations and a smooth PES profile similar to those (Fig.~2) for the moir\'e patterns with smaller unit cells is obtained.

The dependence of the amplitude of PES corrugations on the number $N_c$ of PES unit cells per a graphene unit cell is shown in Fig.~3. This amplitude obtained in the calculations for rigid layers declines nearly exponentially upon decreasing the size of the PES unit cell or equivalently increasing the size of the moir\'e pattern unit cell. The analogous exponential decrease of the amplitude of PES corrugations upon increasing the moir\'e pattern unit cell was found previously for rigid finite graphene layers \cite{Xu2013} for which the rim contribution is the dominant one.

\begin{figure}
   \centering
 \includegraphics{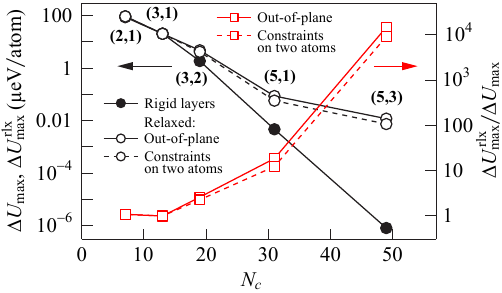}
   \caption{(Color online) Amplitude $\Delta U_\mathrm{max}$ (per atom of the top layer, circles, left axis) of corrugations of the potential energy surface (PES) for interlayer interaction of twisted graphene bilayers and ratio $\Delta U_\mathrm{max}^\mathrm{rlx}/\Delta U_\mathrm{max}$ (squares, right axis) of the amplitudes of PES corrugations for relaxed and rigid layers as functions of the number $N_c$ of PES unit cells per graphene unit cell. 
   The results for rigid layers are shown with filled symbols. The results with structure relaxation with constraints on in-plane positions of all atoms are shown with open symbols and solid lines. The results with structure relaxation with constraints applied to two atoms in the simulation cell are shown with open symbols and dashed lines. Coprime indexes $(n_1,n_2)$ of the commensurate moir\'e patterns are indicated.}
   \label{fig:03}
\end{figure}

The calculations with account of structure relaxation for the both types of atomic constraints considered lead to the same qualitative conclusion that the amplitude of PES corrugations rapidly decreases upon increasing the size of the moir\'e pattern unit cell (Fig.~3, Table~\ref{table:01}). However, the quantitative and even qualitative characteristics of the PES can change significantly as a result of structure relaxation depending on the size of the moir\'e pattern unit cell. For two commensurate moir\'e patterns (2,1) and (3,1) with the small size of the moir\'e pattern unit cell, structure relaxation has a negligible effect on the PES. However, for larger moir\'e pattern unit cells, the influence of structure relaxation is much more prominent and it leads to the increase of the amplitude of PES corrugations (and, correspondingly, to the related increase of the static friction force). This increase grows with the size of the moir\'e pattern unit cell and reaches four orders of magnitude for the moir\'e pattern (5,3) (Fig.~3). Moreover, for the moir\'e patterns (5,1) and (5,3), the structure relaxation causes the change of the PES type from the second type with a trigonal lattice of maxima obtained for rigid layers to the first one with a trigonal lattice of minima. Thus, the calculations performed here show that structure relaxation can lead to a drastic increase of friction in superlubric systems and needs to be taken into account in studies of tribological properties.

The PES amplitudes obtained by relaxation of bilayers with two types of atomic constraints are of the same order of magnitude for all the moir\'e patterns considered (Table~\ref{table:01}). For the moir\'e patterns (2,1) and (3,1) with the small size of the moir\'e pattern unit cell, the relative difference in the PES amplitudes for different types of constraints does not exceed 0.8\%. However, upon increasing the size of the moir\'e pattern unit cell, the relative difference in the PES amplitudes grows and reaches 38\% for the moir\'e pattern (5,3). Still it can be concluded that out-of-plane relaxation provides the major contribution to differences in the PES for relaxed and rigid layers for the considered sizes of the moir\'e pattern unit cell.

We also include in Table~\ref{table:01} the barrier $\Delta U_\mathrm{b}$ to relative sliding of the layers that corresponds to the relative energy of the PES saddle point with respect to the energy minima. This barrier approximately equals $8\Delta U_\mathrm{max}/9$ for the PESs of the first type and $\Delta U_\mathrm{max}/9$ for the PESs of the second type. Similar to the amplitude $\Delta U_\mathrm{max}$ of PES corrugations, the barrier $\Delta U_\mathrm{b}$ decreases fast upon increasing the size of the moir\'e pattern unit cell and changes in the same way as $\Delta U_\mathrm{max}$ upon structure relaxation.

Structural changes associated with structure relaxation have been also analyzed (Table~\ref{table:02}). The amplitude $\Delta d_\mathrm{max}$ of changes in the interlayer distance during the relative in-plane displacement of the layers exponentially decreases upon increasing the size of the moir\'e pattern unit cell, analogously to the amplitude of PES corrugations. On the contrary, the minimum and maximum values of corrugation of the layer plane, $b_\mathrm{min}$ and $b_\mathrm{max}$, grow upon increasing the size of the moir\'e pattern unit cell.  Note that the corrugation ranges are virtually the same for two type of atomic constraints applied during the relaxation. The differences in the optimal interlayer distances $d_\mathrm{min}$ for different constraints are within 0.014\%. 

For small twist angles (i.e.~large moir\'e pattern cells), the relaxed structure of twisted bilayers consisting of identical layers corresponds to commensurate domains separated by incommensurate domain walls \cite{Alden2013, Huang2018, Lebedeva2021} with the width of about 10~nm \cite{Alden2013, Popov2011, Lebedeva2016}. Thus, further increase of the moir\'e pattern cell (or decrease of the twist angle) should lead to the gradual crossover from rather small structural changes related to structure relaxation found here to the system of commensurate domains \cite{Yoo2019}. The size $L$ of the moir\'e pattern unit cell is greater than the the domain wall width of 10~nm for the twist angle $\theta < 1.4^{\circ}$ \cite{Campanera2007}. For moir\'e patterns with a smaller twist angle $\theta$, static friction can be related with the domain wall motion. Such a motion has been observed for graphene/h-BN heterostructure \cite{Mandelli2017} and proposed to occur for graphene layers \cite{Brilliantov2023}. A study of static friction for macroscopic systems of identical layers with a small twist angle (and, correspondingly, a domain wall network) needs a principally different model for atomistic simulations and, therefore, is beyond the scope of the present paper.

Since tribological properties of twisted bilayers consisting of identical layers are determined by the contact size $D$ and twist angle $\theta$,
the ($D$---$\theta$) diagram has been used to show the values of $D$ and $\theta$ corresponding to superlubric and non-superlubric systems \cite{Bai2022a}. The analogous ($D$---$\epsilon$) diagram, where $\epsilon$ is the tensile strain, has been also considered for coaligned graphene bilayer with one stretched layer \cite{Bai2022}. These diagrams were presented for the contact size $D$ up to several tens nanometers for which the rim contribution to friction is dominant. The results on the influence of structure relaxation on static friction obtained here allow us to speculate regarding friction modes of macroscopic superlubric systems corresponding to twisted bilayers consisting of identical layers in the limit $D \rightarrow \infty$ in which the area contribution to static friction is dominant for commensurate moir\'e patterns (Fig.~4). Namely, we assume that for twisted graphene bilayer with the twist angle close to the maximum value $\theta = 30^{\circ}$, static friction should be nearly the same as for rigid layers. For the twist angle $\sim$\,$20^{\circ}$, the first crossover to the friction mode related to the considerable influence of structure relaxation should occur. Upon further decrease of the twist angle, structure relaxation becomes more and more important and for the twist angle $\theta \approx 1.4^{\circ}$, the second crossover to the static friction mode related to the motion of walls between commensurate domains should take place.

\begin{figure}
   \centering
 \includegraphics{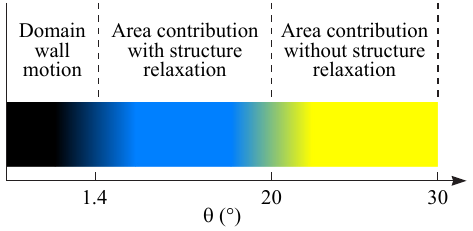}
   \caption{A qualitative diagram of static friction modes of an infinite twisted graphene bilayer with a commensurate moir\'e pattern for different twist angles $\theta$.}
   \label{fig:04}
\end{figure}

\subsection{Approximation of PES by the first Fourier harmonics}

\begin{table*}
\caption{Approximation parameter $U_1$ (in $\mu$eV per atom of the top layer) and relative root-mean-square deviation $\varepsilon$ calculated for commensurate moir\'e patterns $(n_1,n_2)$ with rigid graphene layers, layers relaxed with constraints on in-plane positions of all atoms and only two atoms in the simulation cell for approximation of the potential energy surface using Eq.~(\ref{eq_approx}), as well as parameter $d_1$ (in \AA) and relative root-mean-square deviation $\varepsilon_d$ for approximation of the average interlayer distance change upon structure relaxation using Eq.~(\ref{eq_approx_d}).}
\renewcommand{\arraystretch}{1.2}
\setlength{\tabcolsep}{6pt}
%\resizebox{0.8\textwidth}{!}{
\begin{tabular}{*{11}{c}}
\hline
\hline
 & \multicolumn{2}{c}{rigid layers} & \multicolumn{4}{c}{out-of-plane relaxation} & \multicolumn{4}{c}{relaxation with two atoms constrained} \\
$(n_1,n_2)$ & $U_1$ & $\varepsilon$ &  $U_1$ & $\varepsilon$  & $d_1$ & $\varepsilon_d$ &  $U_1$ & $\varepsilon$  & $d_1$ & $\varepsilon_d$ \\
\hline
(2,1) & $-$19.9 & 1.5$\cdot10^{-2}$ & $-$21.4 & 1.1$\cdot10^{-2}$ & $-$6.08$\cdot10^{-4}$ & 9.6$\cdot10^{-3}$ & $-$21.6 & 2.1$\cdot10^{-2}$ & $-$6.22$\cdot10^{-4}$ & 2.9$\cdot10^{-2}$ \\
(3,1) & 4.58 & 1.0$\cdot10^{-4}$ & 4.52 & 5.5$\cdot10^{-5}$ & 1.17$\cdot10^{-4}$ & 2.1$\cdot10^{-3}$ & 4.49 & 4.4$\cdot10^{-3}$ & 1.16$\cdot10^{-4}$ & 9.2$\cdot10^{-3}$ \\
(3,2) & 0.416 & 5.4$\cdot10^{-4}$ & 1.07 & 2.8$\cdot10^{-4}$ & 3.84$\cdot10^{-5}$ & 8.1$\cdot10^{-3}$ & 0.934 & 1.5$\cdot10^{-3}$ & 3.12$\cdot10^{-5}$ & 6.4$\cdot10^{-3}$ \\
(5,1) & 1.05$\cdot10^{-3}$ & 3.8$\cdot10^{-5}$ & $-$1.93$\cdot10^{-2}$ & 5.4$\cdot10^{-5}$ & $-$6.84$\cdot10^{-7}$ & 2.5$\cdot10^{-1}$ &  $-$1.29$\cdot10^{-2}$ & 1.2$\cdot10^{-3}$ & $-$4.37$\cdot10^{-7}$ & 2.3$\cdot10^{-1}$ \\
(5,3) & N/A\footnote{Not possible to check the adequacy of the approximation because the amplitude is smaller or comparable to the noise in the calculation results.} & N/A & $-$2.66$\cdot10^{-3}$ & 3.2$\cdot10^{-5}$ &  N/A$^\mathrm{a}$ &  N/A & $-$1.66$\cdot10^{-3}$ & 2.4$\cdot10^{-3}$ &  N/A$^\mathrm{a}$ &  N/A \\
\hline
\hline
\end{tabular}
%}
\label{table:prop}
\end{table*}
For a variety of systems consisting of hexagonal layers bound by van der Waals forces, the interlayer interaction PES can be described by the first terms of the Fourier series expansion. Examples of such systems include hexagonal boron nitride (h-BN)  \cite{Lebedev2016, Zhou2015}, graphene \cite{Ershova2010, Lebedeva2011, Popov2012, Zhou2015, Reguzzoni2012}, h-BN/graphene heterostructure \cite{Jung2015, Kumar2015, Lebedev2017}, hydrofluorinated graphene \cite{Lebedev2020} and double-layer graphene with an inert gas spacer \cite{Popov13b}. Recently we found that the PES of interlayer interaction energy for rigid layers forming a commensurate moir\'e pattern can be also approximated by the first Fourier harmonics \cite{Minkin2023}. Here we show that in spite of the drastic influence of structure relaxation on the amplitude of PES corrugations for latter systems, this approximation is valid with a high accuracy not only for rigid but also for relaxed structure of the layers.

As discussed in \cite{Minkin2023}, for twisted honeycomb lattices, the PES is periodic with respect to translation along a lattice vector of either of the two layers. Thus, only reciprocal vectors representing overlapping vertices of reciprocal lattices of the layers contribute to the PES Fourier transform. The reciprocal lattices of twisted honeycomb layers create a moir\'e pattern in the same way as real-space lattices of the layers create the moir\'e pattern in real space. The contribution of the first Fourier harmonics to the PES, therefore, looks like \cite{Minkin2023}
\begin{equation} \label{eq_approx}
\delta U(x',y') = U_1\bigg(2\cos{(k'_yy')}\cos{(k'_xx')} +\cos{(2k'_yy')}\bigg).
\end{equation}
Here $x'$ axis is aligned along one of the moir\'e pattern vectors, $y'$ axis is aligned in the perpendicular direction ($x' = x \cos\varphi - y \sin\varphi$,  $y' = y \cos\varphi + x \sin\varphi$), $k'_x = \sqrt{N_c}k_x$ and $k'_y = \sqrt{N_c}k_y$.

To check how well Eq.~(\ref{eq_approx}) describes the PES, we have approximated the computed dependences of the potential energy on the displacement along the segment AB directed along the $y'$ axis for bilayers with rigid layers and layers relaxed with constraints applied to all or two atoms of the simulation box. The parameter $U_1$ in Eq.~(\ref{eq_approx}) obtained by minimization of the root-mean-square deviation and the relative deviation $\varepsilon$, which is given by the root-mean-square deviation divided by $\Delta U_\mathrm{max}$, are listed in Table~\ref{table:prop}. Negative values of the parameter $U_1$ correspond to the first PES type (Fig.~2a), whereas positive values to the second one (Fig.~2b).

It is seen from Table~\ref{table:prop} that the PES is approximated well by Eq.~(\ref{eq_approx}) both for rigid and relaxed structures of the layers. In most cases, the relative deviation decreases when taking into account out-of-plane relaxation but increases when in-plane atomic displacements are considered. This might be related to the deficiency of how the relative displacement of the layers is induced in the latter case. The fact that atoms are not constrained in the same manner in the moir\'e unit cells within the simulation box can give rise to non-local phenomena and slightly different behavior of the moir\'e unit cells. Nevertheless, even in that case, the relative deviation for Eq.~(\ref{eq_approx}) barely exceeds 2\%.

It is interesting that structure relaxation maintains the PES shape. Previously we approximated the PESs for these commensurate moir\'e patterns with rigid structure of the layers on a grid of points within the whole PES unit cell \cite{Minkin2023}. Those approximations gave nearly the same values of the parameter $U_1$ with the difference from the values calculated here within 1\%. This confirms that the approximation along the segment AB performed here is an adequate way to reproduce the whole approximated PES. Note that in studies of other 2D materials, the relative deviations of the approximated PESs were about 3\% for hydrofluorinated graphene \cite{Lebedev2020}, 1\% for graphene \cite{Popov2012, Lebedeva2011}, 0.3\% for graphene/h-BN heterostructure \cite{Lebedev2017} and 0.1--0.3\% for h-BN \cite{Lebedev2016}.

Based on Eq.~(\ref{eq_approx}) for the approximated PESs, the expressions for a set of physical quantities related with interlayer interaction and including the barrier for relative rotation of commensurate twisted layers to an incommensurate state,  shear strength, shear modulus and shear mode frequency were obtained in our previous work \cite{Minkin2023}. All these quantities are determined by a single energy parameter of the PES. Particularly, the increase in the amplitude of PES corrugations because of structure relaxation means the proportional increase of the barrier mentioned above. It was proposed that the presence of such a barrier can lead to robust macroscopic superlubricity in systems of commensurate twisted layers \cite{Minkin2023}. Therefore, structure relaxation should favor this mechanism of robust macroscopic superlubricity.

As for the static friction force, measured values of this force depend on the path on the PES during relative motion of layers in an experiment. Complex atomistic models describing a particular experimental set-up are necessary to properly compute it (see, for example, \cite{Verhoeven2004, Dienwiebel2004, Filippov2008} for motion of a graphene flake attached to a probe on graphite surface). Nevertheless, the static friction force $f_\mathrm{s}$ per atom of one of the layers for motion along the minimum energy path (a zigzag line of segments connecting neighboring minimums through saddle points) can be found as $f_\mathrm{s} = 4\pi\sqrt{N_c}U_1/a$ for the PES of the first type and $f_\mathrm{s} = 2.677\sqrt{N_c}U_1/a$ for the second PES type \cite{Minkin2023}. These expressions demonstrate that such a force is proportional to the amplitude of PES corrugations.

The studies of the effect of structure relaxation on the PES for coaligned graphene and h-BN layers in different combinations \cite{Zhou2015} showed that the dependence of the  optimized interlayer distance on the displacement can also be described by the first Fourier harmonics. Therefore, we approximate the change of the average interlayer distance upon structure relaxation by the expression similar to Eq.~(\ref{eq_approx}) with the parameter $d_1$ instead of $U_1$
\begin{equation} \label{eq_approx_d}
\delta d(x',y') = d_1\bigg(2\cos{(k'_yy')}\cos{(k'_xx')} +\cos{(2k'_yy')}\bigg).
\end{equation}

The parameter $d_1$ obtained by minimization of the root-mean-square deviation and the relative root-mean-square deviation $\varepsilon_d$ are also given in Table~\ref{table:prop}. It is seen that the variation in the interlayer distance is described well by the first Fourier harmonics, especially for moir\'e patterns with small unit cells. However, the relative deviation $\varepsilon_d$ grows with increasing of the size of the moir\'e pattern unit cell. This can be related with the following reasons. First, the amplitude of interlayer distance variation decreases strongly upon increasing the size of the moir\'e pattern unit cell. Because of the noise in the interlayer distance variation computed using the interatomic potential of the magnitude of about $10^{-6}$ \AA, the relative accuracy of the calculations becomes worse. Second, with the decrease of the twist angle (i.e.~with the increase of the size of the moir\'e pattern unit cell), the relaxed structure of twisted graphene bilayer gradually transforms  into the system of commensurate domains separated by incommensurate domains walls \cite{Lebedeva2021}. Both of these reasons lead to the increase of the relative deviation $\varepsilon_d$.

\section{Discussion and conclusions}

The restriction of macroscopic structural superlubricity due to atomic structure relaxation has been studied by the example of infinite twisted graphene bilayers. For this purpose, the potential energy surfaces (PESs) for in-plane relative displacements of the layers have been calculated for twisted graphene bilayers with commensurate moir\'e patterns (2,1), (3,1), (3,2), (5,1), and (5,3) considering periodic boundary conditions and using the classical registry-dependent Kolmogorov--Crespi potential with and without account of structure relaxation.

Two types of constraints have been considered to control the relative in-plane displacement of the layers. In the first case, in-plane positions of all atoms of the layers are fixed and relaxation is limited to the out-of-plane degree of freedom. In the second case, such constraints are applied only to two atoms of the supercell, while the rest of the atoms are completely free. Our calculations  show that the results obtained with these two types of constraints are very close for the moir\'e patterns (2,1) and (3,1) with the small unit cells but the difference grows with the moir\'e pattern unit cell. Still even for moir\'e patterns with large unit cells, the differences in the amplitude of PES corrugations are within the factor of 2. They are much smaller than the changes with respect to rigid layers, which correspond to orders of magnitude. Therefore, out-of-plane relaxation is sufficient to capture the most important relaxation effects. In experiments, the behavior of a system of stacked layers upon their relative motion is determined by forces applied to the layers. The forces can be applied equally to all atoms (e.g., in the case of an accelerometer \cite{Kim13}), to layer edges, to a small region of a layer using a probe tip, etc. A simulation of system behavior under experimental conditions requires the use of complex models and usually limits the consideration of relative motion of the layers only to particular sliding paths close to the PES minima and saddle points, while the rest of the PES is not investigated. By considering two types of constraints described above, we have obtained the whole PES describing the potential energy dependence on the local stacking. Such a PES can be employed in large-scale models, e.g. Frenkel--Kontorova model \cite{Popov2011, Lebedeva2019, Lebedeva2016}, continuum methods \cite{Yoo2019, Zhang2018a}, etc., for simulations of superstructures and macroscopic phenomena in twisted graphene layers.

It is found that the amplitude of PES corrugations (that is the difference between the minimum and maximum values of the PES) calculated with and without the account of structure relaxation rapidly decreases upon increasing the size of the moir\'e pattern unit cell. This is similar to what was obtained in the calculations with rigid layers for finite twisted graphene bilayers with commensurate moir\'e patterns \cite{Xu2013}, where the rim contribution into static friction was dominant. The influence of structure relaxation on the amplitude of PES corrugations strongly depends on the unit cell size of the moir\'e pattern. For the moir\'e patterns (2,1) and (3,1) with the smallest unit cells, the amplitudes of PES corrugations calculated with and without account of the structure relaxation are approximately the same. However, for the moir\'e patterns (3,2), (5,1), and (5,3) with greater unit cells, the structure relaxation leads to an increase of the amplitude of PES corrugations. This increase becomes greater with increasing the unit cell size and reaches four orders of magnitude for the moir\'e pattern (5,3). The structure relaxation can even affect qualitative characteristics of the PES. Namely, for the moir\'e patterns (5,1) and (5,3), the structure relaxation causes the change of the PES type from the second type with a trigonal lattice of maxima obtained for rigid layers to the first one with a trigonal lattice of minima.

We have demonstrated that the PESs calculated with and without account of structure relaxation can be approximated by the first Fourier harmonics determined by symmetry of the commensurate moir\'e pattern with an accuracy within 2\% relative to the amplitude of PES corrugations. Additionally it is shown that corrugations of the interlayer distance for the relaxed structure can be approximated by the analogous expression which contains only the first Fourier harmonics. The approximation by the first Fourier harmonics for coaligned layers was used previously for the analysis of electronic properties of graphene/h-BN heterostructure \cite{Jung2014, Jung2015} and twisted graphene \cite{Jung2014}. We expect that expressions similar to the ones used here for the PES and interlayer distance can also be used to describe electronic properties of 2D systems with relaxed twisted layers.

Since the static friction force is directly determined by the PES for in-plane displacements of the layers (which has a simple shape determined by the first Fourier harmonics), our results on the influence of structure relaxation on the amplitude of PES corrugations are equally valid for the static friction force. Therefore, we assume that for a macroscopic superlubric system consisting of identical layers with the twist angle close to the maximum value, static friction is nearly the same as for rigid layers. Upon decreasing the twist angle, the first crossover to the friction mode related to the considerable influence of structure relaxation occurs. With the further decrease of the twist angle, the structure relaxation leads to formation of commensurate domains and the second crossover to the static friction mode related to the motion of domain walls takes place. Dynamic friction should also correlate with the amplitude of PES corrugations due to the dissipation of the kinetic energy of relative motion of the layers on PES hills \cite{Popov2011a}. Thus, we believe that the account of structure relaxation of both layers is important for consideration of dynamic friction in superlubric systems as well.

\section*{Acknowledgments}

A.S.M., A.M.P. and Y.E.L. acknowledge the support by the Russian Science Foundation grant No. 23-42-10010, https://rscf.ru/en/project/23-42-10010/, for the results described in Sec.~IIIA ``Influence of structure relaxation''. I.V.L. acknowledges the IKUR HPC project ``First-principles simulations of complex condensed matter in exascale computers'' funded by MCIN and by the European Union NextGenerationEU/PRTR-C17.I1, as well as by the Department of Education of the Basque Government through the collaboration agreement with nanoGUNE within the framework of the IKUR Strategy. A.M.P. and Y.E.L. acknowledge the support by project FFUU-2024-0003 of the Institute of Spectroscopy of the Russian Academy of Sciences for the results described in Sec.~IIIB ``Approximation of PES by the first Fourier harmonics''. S.A.V. and N.A.P. acknowledge the support by the Belarusian Republican Foundation for Fundamental Research (Grant No.~F23RNF-049) and by the Belarusian National Research Program ``Convergence-2025''. This work has been particularly carried out using computing resources of the federal collective usage center Complex for Simulation and Data Processing for Mega-science Facilities at NRC ``Kurchatov Institute'', http://ckp.nrcki.ru.

The authors declare no conflict of interest.

\section*{Data availability}
The data that support the findings of this article are openly available \cite{Minkin2024}.

\bibliography{PRM2024-moire}
\end{document}